\documentclass[prc,twocolumn,tightenlines,preprintnumbers]{revtex4}
\usepackage{bm}
\usepackage{amssymb}
\usepackage{amsmath}
\usepackage{graphicx}

\begin{document}

\title{The Breaking Strain of Neutron Star Crust and Gravitational Waves}
\author{C. J. Horowitz}\email{horowit@indiana.edu} 
\affiliation{Department of Physics and Nuclear Theory Center,
             Indiana University, Bloomington, IN 47405, U.S.A}
\author{Kai Kadau}\email{kkadau@lanl.gov}
\affiliation{Los Alamos National Laboratory,
Physics and Chemistry of Materials,
Group T-1, MS G756,
Los Alamos, NM 87545, U.S.A}
\date{\today}
\begin{abstract}
Mountains on rapidly rotating neutron stars efficiently radiate gravitational waves.  The maximum possible size of these mountains depends on  the breaking strain of neutron star crust.  With multi-million ion molecular dynamics simulations of Coulomb solids representing the crust, we show that the breaking strain of pure single crystals is very large and that impurities, defects, and grain boundaries only modestly reduce the breaking strain to around 0.1.  Due to the collective behavior of the ions during failure found in our simulations, the neutron star crust is likely very strong and can support mountains large enough so that their gravitational wave radiation could limit the spin periods of some stars and might be detectable in large scale interferometers.  Furthermore, our microscopic modeling of neutron star crust material can help analyze mechanisms relevant in magnetar giant and micro flares. 
\end{abstract}
\smallskip
\maketitle

``Mountains'' on rapidly rotating neutron stars (NS) efficiently radiate gravitational waves (GW) because of their large masses and high accelerations\cite{crustmonster}.  Searches for these waves are presently underway with large scale interferometers\cite{ligo,ligo2}.  Furthermore, the angular momentum radiated could control the spin periods of some accreting NS\cite{watts2008}.  As material falls on a NS, from its binary companion, the angular momentum accreted spins up the star.   However, GW radiation increases very rapidly with rotational frequency.  The star may reach spin equilibrium where the angular momentum gained from accretion is balanced by that lost to GW radiation.  This could explain why the most rapidly spinning pulsars observed are only spinning at about half of the breakup rate.     

How high can a mountain be, before it collapses under the NS's extreme gravity?  This depends on the breaking strain (BS)\cite{footnote1} of NS crust (NSC) which is the fractional deformation when the crust fails.   Estimates of the breaking strain vary by orders of magnitude and are based on lose analogies with conventional materials rather than on detailed calculations or simulations\cite{breakingestimate1,breakingestimate2}.  This is, perhaps, the largest uncertainty in estimating upper limits for these GW.  In this paper we use molecular dynamics (MD) simulations to significantly improve estimates of the breaking strain.      
   
Crust breaking may also be important for magnetar giant flares.  These are extremely energetic Gamma-ray bursts from strongly magnetized NS\cite{giantflares}.  Thompson and Duncan model giant flares as ``star quakes"\cite{duncan}.  A strong twisted magnetic field stresses the crust until it breaks.  The crust then moves and allows magnetic field lines to reconnect.  This releases magnetic energy that powers the flare.   This mechanism may require the crust to be relatively strong in order for it to control the very strong magnetic field.

How the crust breaks may be important for the excitation of NS oscillation modes from a ``star quake".  Quasiperiodic oscillations have been observed in the tails of giant flares\cite{qpo1,qpo2}.  These have been interpreted as shear oscillations of the crust.  If these modes can be convincingly identified, they may provide considerable information on the structure of the NS and its crust\cite{QPOinterp3}.  In addition, LIGO (Laser Interferometer Gravitational Wave Observatory) has searched for GW from a magnetar giant flare that could come from large amplitude NS oscillations\cite{GWflares}.  Our simulations of crust breaking may lead to insight into the amplitudes that might be expected.  

In the crust of a NS, electrons form a very degenerate relativistic gas.  The ions are completely pressure ionized and have Coulomb interactions that are screened at large distances by the slightly polarizable electron gas.  The interaction between two ions $i,j$ is assumed to be a screened Coulomb or Yukawa potential\cite{screening},
\begin{equation}
\phi(r_{ij})=\frac{Z^2 e^2}{r_{ij}}{\rm e}^{-r_{ij}/\lambda_e}\, ,
\end{equation}
where the ions have charge $Z$, $r_{ij}$ is the distance between them, and the electron screening length $\lambda_e=\pi^{1/2}/2e(3\pi^2n_e)^{1/3}$ with $n_e$ the electron density.  The total potential energy is given by the sum over all pairs $\sum_{i<j}\phi(r_{ij})$. Charge neutrality insures that $n_e=Zn$ where $n$ is the ion density.  The ions are assumed to form a classical one component plasma that can be characterized by the Coulomb parameter $\Gamma$,
\begin{equation}
\Gamma=\frac{Z^2e^2}{a T}\, .
\label{gamma}
\end{equation}
This is the ratio of a typical Coulomb energy to thermal energy and the ion sphere radius $a=[3/(4\pi n)]^{1/3}$ is a typical distance between ions.

We calculate at a reference density of $n=7.18\times 10^{-5}$ fm$^{-3}$ using $Z=29.4$ and an atomic mass number $A=88$ ($1\times 10^{13}$ g/cm$^3$)\cite{phasesep}.  Alternatively, nuclei are very neutron rich near the base of the crust.  For $A=880$ this $n$ corresponds to $10^{14}$ g/cm$^3$.  Results can be scaled to other densities at constant $\Gamma$ and approximately scaled to other $Z$ at constant $\Gamma$,  which involves only a small change in $\lambda_e$.  Most of our simulations are for a temperature $T=0.1$ MeV.  This corresponds to $\Gamma=834$.  We ignore any strong interactions between ions and the effects of free neutrons that are present in the inner crust.

\begin{figure}[htp]
\centering
\includegraphics[width=4in,angle=0,clip=true] {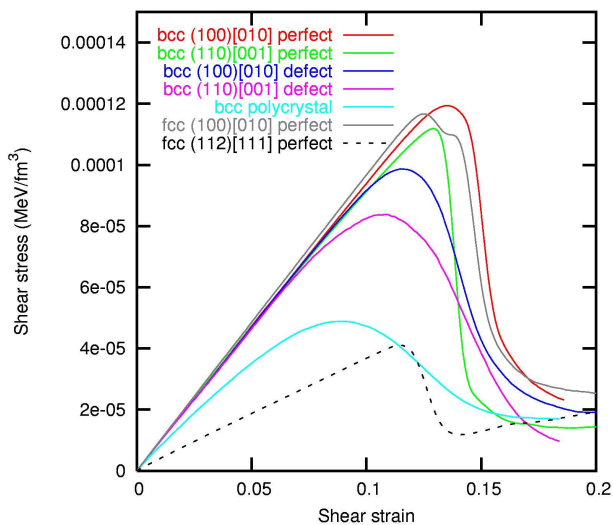}
\caption{Shear stress versus strain for perfect and defective bcc and fcc single crystals containig about 2 million ions for different shear systems ---as given by the crystallographic orientation of the shear plane and the direction of the applied stress \cite{baskes} ---  are shown. In addition a polycrystalline sample  containing 12.8 million ions and 8 randomly oriented grains with an average grain diameter of 3962~fm is shown.  Results were obtained at a strain rate of 4$\times 10^{-7}$ c/fm.}
\label{Fig1a}
\end{figure}

We perform large-scale MD simulations of shearing NSC material as we believe that this is the most important mode of failure in the crust, although real neutron star strain fields could involve some combination of shear strain and tension.  Tension simulations at constant volume, had a breaking strain and strength that was smaller by 2.5 and 2, respectively.  We use two independent MD codes: YukawaMD is a serial code where the system is strained by deforming the periodic boundary.   An originally cubic simulation volume is deformed according to $x\rightarrow x+\epsilon y/2$, $y\rightarrow y+\epsilon x/2$, and $z\rightarrow (1+\epsilon^2/2)z$.   Here the strain $\epsilon$ is increased linearly with time.

SPaSM\cite{spasm} is a high-performance parallel MD code where the system is strained by moving a boundary layer of frozen ions on top and bottom against each other\cite{baskes}.  In order to simulate systems containing up to 12.8 million ions we employed up to 512 processors of a parallel architecture consisting of AMD Opteron cores @ 2.2 GHz installed at Los Alamos (Yellowrail). The computational cost for these simulations are about 100 times more as compared to short range Lennard-Jones computations because of the much longer ranged force ($r_{\rm c}=10\lambda_e$).  The shear is initiated by having a frozen layer of ions, i.e. ions that are not dynamic, on the top and bottom of the system. The layers are separated by the distance $d$ and moved with a velocity $v$ against each other to initiate the desired shear rate  $d\epsilon/dt= v/d$.
We cutoff the potential at the large distance  $r_{\rm c}=10\lambda_e$\cite{footnote4} resulting in each ion interacting with about 5400 neighbors; for smaller values of $r_{\rm c}$ the results depend on $r_{\rm c}$\cite{shearmod}.  See supplemental Figures (S1,S2) \cite{supplement}.

Figure \ref{Fig1a} shows the shear stress versus strain for body-centered cubic (bcc) crystals, as this is the equilibrium structure.  Some results for the face-centered cubic (fcc) structures are added for comparison.  For all investigated crystallographic shear systems the perfect crystals show a BS well above 0.1, and break in a rather abrupt fashion with only a very small region where plasticity, i.e. deviation from a linear stress-strain relation, is present.  The multi-million ion systems were strained at a rate of 4$\times 10^{-7}$ c/fm. As we cannot simulate the very large time and length scales associated with NSC we have to rely on estimates based on a series of simulations that suggested basically no size effects for the single crystal simulations presented here (Fig. S3 \cite{supplement}) and a converging result at low shear rates (Fig. S4 \cite{supplement}). We also note that tripling the temperature only reduces the maximum stress by about 25\% and does not significantly alter the BS.

\begin{figure}[htp]
\centering
\includegraphics[width=3.2in,angle=0,clip=true] {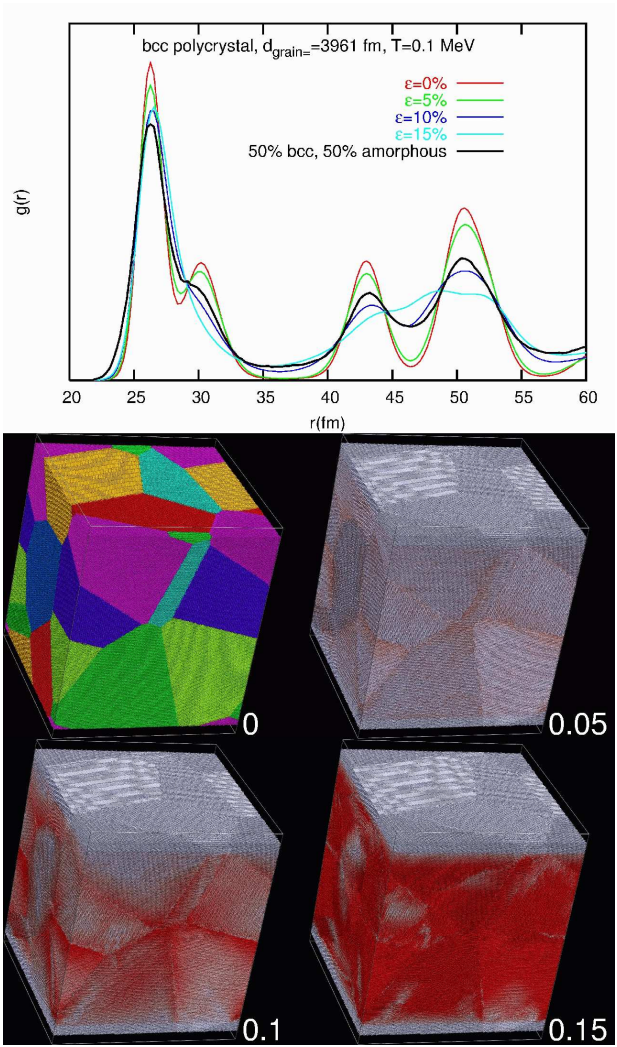}
\caption{Polycrystalline sample with 12.8 million ions consisting of 8 differently oriented grains with an average grain diameter of 3961 fm  at strains of 0.0, 0.05, 0.1, and 0.15.  Upper panel: The radial distribution functions at different shear states exhibit the characteristic peaks of the bcc structure for strains up to 0.1. At 0.15 strain some degree of amorphization might be possible. Lower panel: At 0.0 strain the eight different grains are shown in different colors. At strains of 0.05, 0.1, and 0.15 the red color indicates plastic deformation, i.e. deviations of the ions away from an ideally uniformly sheared bcc lattice in the range of 0 to two times the nearest neighbor distance. Note that the top and bottom layers are frozen and moved to impose the shear.}
\label{Fig2}
\end{figure}    
 
As very little is known about the defect structure, grain sizes, and associated grain boundaries of NSC that potentially can reduce the strength, we also consider polycrystalline materials that are generated using a Voronoi construction in which grain centers and orientations are picked at random and each ion belongs to the grain center that is closest\cite{voronoi}.  After this initialization, the system is equilibrated by heating it from 0.1 to 0.3 MeV and back to 0.1 MeV over a total simulation time of 275000 fm/c.  During equilibration at elevated temperatures, ions find lower energy configurations exponentially faster, relaxing the initial artificial grain boundary structures more rapidly.   The system is then sheared at a strain rate of $4\times 10^{-7}$ c/fm.  Figure 2 shows one system with 12.8 million ions at strains of 0.0, 0.05, 0.1, and 0.15 (for stress-strain curve see Fig.1).  The radial distribution functions exhibit characteristic peaks of the bcc structure for strains up to 0.1. After failure, at 0.15 strain some degree of amorphization might be possible. The system starts to deform plastically near the grain boundaries in a collective manner without exhibiting signs of dislocations or other more localized events (Fig.2, See also supplemental Movies S5,S6 \cite{supplement}). However, at large strains the samples are plastically deformed and the mode of failure seems to be a reorientation of local regions ---possibly associated with some degree of amorphization (Fig.2)---  to accommodate the shear. This collective, rather than more localized dislocation-based, mode of plasticity makes the crystal stronger and break at larger strains than  terrestrial metals, where the electronic density can have a localized structure to accommodate local defects. This failure mechanism also does not allow for voids or fracture to appear as these localized defects would possibly heal under the influence of the high pressure. In fact, simulations that started out with a cylindrical hole with a diameter of 2.5$\times$ the nearest neighbor distance initialized into the otherwise perfect single crystal, showed the void quickly heals under the influence of the enormous pressure present in the system and the maximum stress is only  reduced by 25\% (Fig.1, Movie S7 \cite{supplement}). This is consistent with the prediction of Jones that voids would not form because of the high pressure\cite{jones}. Furthermore, simulations containing only two grains and no frozen-in ions rapidly evolved into a single crystal\cite{thermalconductivity}. We investigated samples with grain diameters between 570 fm and 3961 fm that showed a clear monotonic trend to increase the BS and stress with the grain size. As we expect to have larger grains in NSC material than we could possibly simulate on a computer we expect the crust material strength only to be slightly reduced by the presence of grain boundaries. Also, note that accreting NS can form new crust very slowly over accretion times of thousands of years.  This should allow plenty of time for larger crystals to form.  Second, samples with very small grain size may have a reduced thermal conductivity and this could disagree with crust cooling observations\cite{shternin,cumming}.   Indeed, the grain size could be much larger than in our simulations.  In that case, we expect the strength of the system to be larger and closer to our single crystal results.

To study the role of impurities we form a 27648 ion crystal made of the complex rapid proton capture nucleosynthesis ash considered in ref.\cite{phasesep}.  We use the solid composition from Table I of ref.\cite{phasesep}.  This includes a range of ions from $Z=8$ to 47 and can be characterized by an impurity parameter $Q=\langle Z^2 \rangle -\langle Z \rangle^2=22.3$ that describes the dispersion in ion charges.   Shternin et al.\cite{shternin} and Brown and Cumming \cite{cumming} study the rapid crust cooling of two NS, after extended outbursts.  Brown and Cumming conclude that Q must be less than 10\cite{cumming}, see also\cite{thermalconductivity}.  Otherwise impurities would reduce the crust thermal conductivity by too much to agree with observation.  Our crystal has more impurities than this limit.  Therefore, we believe our results may represent an upper limit on the effects of impurities.  Indeed, we have so many impurities that our runs with significantly more impurities phase separate\cite{phasesep}.  We equilibrate our crystal by heating it up and then slowly cooling it back down to $T=0.1$ MeV over a total simulation time of about 85 million fm/c.  We shear this crystal at a strain rate of $8\times 10^{-7}$ c/fm and find a maximum stress that is reduced from the maximum stress of a pure crystal by less than 45\% at a similar strain.  We conclude that impurities are unlikely to significantly reduce the strength of NSC.

Our simulations show that the breaking strength of neutron star crust is about $10^{10}$ times more than for terrestrial engineering materials such as metal-alloys where the strength is measured in fractions of a GPa. The largest contributor to this tremendous difference is of course the enormous pressure and thus density of the crust.  Furthermore, the screened Coulomb interaction is purely repulsive and has no explicit length scale\cite{footnote5}, i.e. the system at twice the density behaves just like the system at the original density only at a lower temperature (Eq.\ref{gamma}).  This causes the material to fail abruptly in a collective manner at a large strain, rather than yielding continuously at low strain as observed in metals, because of the formation of dislocations.  For example, the breaking strain of steel is around 0.005, some twenty times smaller than what we find for the neutron star crust.  We speculate that the collective plastic behavior found here could help to improve design strategies that suppress the weakening effects of dislocations and other more localized defects in conventional materials.  Note that small Coulomb solids have been studied in the laboratory using cold trapped ions\cite{coldatoms}.

Gravitational wave radiation depends on the ellipticity $\epsilon$ (fractional difference in moments of inertia) of a rotating star.   We estimate, following ref. \cite{crustmonster},  that our breaking strain $\approx 0.1$ can support an $\epsilon\le 4\times 10^{-6}$ for a 1.4 solar mass, 10 km radius NS.   This ellipticity, for a rapidly rotating star, will generate GW that could be detectable by LIGO \cite{ligo2}.

In conclusion, we have performed large-scale MD simulations of the breaking strain (BS) of Coulomb solids representing neutron star crust (NSC).  We find a collective mode of failure that does not involve localized defects such as dislocations, opening of voids, or fractures.  The breaking strain in the presence of introduced defects and impurities is only moderately reduced to about 0.1.  The large breaking strain, that we find, should support mountains on rapidly rotating NS large enough to efficiently radiate gravitational waves.  This motivates further work on mountain forming mechanisms and searches for continuous gravitational waves.  Furthermore, the methods we presented here are very promising to describe other aspects of Coulomb solids and neutron star crust in particular.

 We thank Ben Owen, Timothy C. Germann, Sanjay Reddy, and Bob Rutledge for important discussions.  We also thank Andrey Chugunov for comments on the manuscript and Don Berry for help in equilibrating the impure sample.  This work was supported in part (C.H.) by DOE grant DE-FG02-87ER40365 and by Shared University Research grants from IBM, Inc. to Indiana University and partly (K.K.) under the auspices of the National Nuclear Security Administration of the U.S. Department of Energy at Los Alamos National Laboratory under Contract No. DE-AC52-06NA25396 with funding from the LDRD-DR project {\it X-ray bursts, Superbursts and Giant Flares}.


\begin{thebibliography}{99} 
\bibitem{crustmonster} G. Ushomirsky, C. Cutler, L. Bildsten, {\it Mon. Not. R. Astron. Soc.} {\bf 319}, 902 (2000).
\bibitem{ligo}B. Abbott {\it et al}., {\it Astrophys. J.} {\bf 683}, L45 (2008).
\bibitem{ligo2} B. Abbott {\it et al}., {\it Phys. Rev. D} {\bf 76}, 042001 (2007).
\bibitem{watts2008} A. L. Watts, B. Krishnan, L. Bildsten, B. F. Schutz, {\it Mon. Not. R. Astron. Soc.} {\bf 389}, 839 (2008).
\bibitem{footnote1} In most of the gravitational wave literature the breaking strain is defined as the maximum stress over the shear modulus.  As the deviation from a linear stress-strain relation is only modest before the yield this is very similar to the strain at which the NSC yields.
\bibitem{breakingestimate1} R. Simoluchowski, {\it Phys. Rev. Lett.} {\bf 24}, 923 (1970).  
\bibitem{breakingestimate2} M. Ruderman, {\it Astrophys. J.} {\bf 382}, 587 (1991).
\bibitem{giantflares}K. Hurley {\it et al}., {\it Nature} {\bf 434}, 1098 (2005).
\bibitem{duncan} C. Thompson, R. C. Duncan, {\it Astrophys. J.} {\bf 561}, 980 (2001).
\bibitem{qpo1}G. Isreal {\it et al}., {\it Astrophys. J.} {\bf 628}, L53 (2005).
\bibitem{qpo2} Tod E. Strohmayer, Anna L. Watts, {\it Astrophys. J.} {\bf 632}, L111 (2005).
\bibitem{QPOinterp3} Lars Samuelsson, Nils Andersson, {\it Mon. Not. Roy. Astron. Soc.} {\bf 374}, 256 (2007). 
\bibitem{GWflares}B. Abbott {\it et al}., {\it Phys. Rev. Lett.} {\bf 101}, 211102 (2008).
\bibitem{screening}  A. L. Fetter, J. D. Walecka, {\it Quantum Theory of Many Body Systems} (McGraw-Hill, N. Y.,1971), p. 175.
\bibitem{phasesep} C. J. Horowitz, D. K. Berry, E. F. Brown, {\it Phys. Rev. E} {\bf 75}, 066101 (2007).
\bibitem{spasm}P. S. Lomdahl, P. Tamayo, N. Gronbech-jensen, D. M. Beazley, in {\it Proceedings of Supercomputing 93}, G. S. Ansell, Ed. (IEEE Computer Society Press, Los Alamitos, CA, 1993), p. 520.
\bibitem{baskes}M. F. Horstemeyer, M. I. Baskes, and S.J. Plimpton, {\it Acta Mater.} {\bf 49}, 4363 (2001).
\bibitem{footnote4} In order to have a smooth potential $\phi(r_{\rm c})=0$, in code YukawaMD we modify the potential as follows $\phi(r)\rightarrow [\phi(r)-\phi(r_{\rm c})]\Theta(r_{\rm c}-r)$ ($\Theta$ being the step function).  In the code SPaSM we in addition impose $d\phi(r_{\rm c})/dr=0$, by replacing $\phi(r)\rightarrow \phi(r) (1-(r/r_{\rm c})^{10})^2$. For large cut off distances no difference in the results can be found.
\bibitem{shearmod} C. J. Horowitz, J. Hughto (available at http://arXiv.org/abs/0812.2650).
\bibitem{supplement} Movies S5-S7 are  available at http://cecelia.physics.indiana.edu/breaking-strain 
See also EPAPS Document No. [number will be provided by publisher later] for supplemental Figs S1-S4 and movies S5-S7. For more information on EPAPS, see http://www.aip.org/pubservs/epaps.html.  
\bibitem{voronoi}K. Kadau, T. C. Germann, P. S. Lomdahl, R.C. Albers, J.S. Wark, A. Higginbotham, 
Brad Lee Holian, {\it Phys. Rev. Lett.} {\bf 98}, 135701 (2007).
\bibitem{jones}P. B. Jones, {\it Astrophys. J.} {\bf 595}, 342 (2003).
\bibitem{thermalconductivity}C. J. Horowitz, O. L. Caballero, D. K. Berry, {\it Phys. Rev. E} {\bf 79}, 026103 (2009). 
\bibitem{cumming}  E. F. Brown, Andrew Cumming (available at http://arXiv.org/abs/0901.3115).
\bibitem{shternin} P. S. Shternin, D. G. Yakovlev, P. Haensel, A. Y. Potekhin (available at http://arxiv.org/abs/0708.0086).
\bibitem{footnote5} This is of course only true if the interaction is cut off at a large distance as we do, otherwise for shorter cut off distances the strength can be significantly reduced (Fig.S2).
\bibitem{coldatoms} T. B. Mitchell et al., {\it Phys. Rev. Lett.} {\bf 87}, 183001 (2001).
\end{thebibliography}
\end{document}